\documentclass[preprint2]{aastex}

\shorttitle{Numerical Modeling of Multi-wavelength Spectra of M87 Core Emission}
\shortauthors{Hilburn and Liang}

\begin{document}


\title{Numerical Modeling of Multi-wavelength Spectra of M87 Core Emission}


\author{G. Hilburn and E. P. Liang}
\affil{Physics and Astronomy Department, Rice University,
    Houston, TX 77005}
\email{guy.l.hilburn@rice.edu}



\begin{abstract}
	Spectral fits to M87 core data from radio to hard x-ray are generated via a specially selected software suite, comprised of the HARM GRMHD accretion disk model and a 2D Monte Carlo radiation transport code.  By determining appropriate parameter changes necessary to fit x-ray quiescent and flaring behavior of M87's core, we assess the reasonableness of various flaring mechanisms.  This shows that an accretion disk model of M87's core out to 28 $GM/{c^2}$ can describe the inner emissions.  High spin rates show GRMHD-driven polar outflow generation, without citing an external jet model.  Our results favor accretion rate changes as the dominant mechanism of x-ray flux and index changes, with variations in density of approximately 20\% necessary to scale between the average x-ray spectrum and flaring or quiescent spectra.  The best fit parameters are black hole spin a/M $>$ 0.8 and maximum accretion flow density $n \leq 3\times10^7$ cm$^{-3}$, equivalent to horizon accretion rates between $\dot m = \dot M/\dot M_{Edd} \approx 2\times10^{-6}$ and $1\times10^{-5}$ (with $\dot M_{Edd}$ defined assuming a radiative efficiency $\eta = 0.1$).  These results demonstrate that the immediate surroundings of M87's core are appropriate to explain observed x-ray variability.
\end{abstract}





\section{Introduction}

	It is generally accepted that the center of the Faranoff-Riley type I (FR-I) radio galaxy M87 harbors a supermassive black hole of mass $(6\pm0.5)\times10^9 M_o$ (Gebhardt and Thomas, 2009) at a distance of 16.7 Mpc \citep{mei07}, which is associated with a spectacular kiloparsec scale jet.  Observations of superluminal motion in the jet require a jet viewing angle of $\theta < 19$ degrees and bulk Lorentz factor $\gamma > 6$ at the prominent HST-1 jet knot, which would imply it is located $5.3\times10^5 R_s$ downstream from the core \citep{bir99}, where the Schwarzschild radius of the black hole $R_s = 1.8\times10^{15}$ cm \citep{har10}).  Due to its size, proximity, and orientation, M87 provides a unique opportunity for study of a central AGN environment, which can be probed to investigate particle energization in accretion disks, jet launching, and other astrophysical phenomena occurring in these extreme situations.

	Its spectral energy distribution (SED) suggests that M87 is a misaligned BL Lac.  It has been observed for a number of years from radio to gamma rays, and detailed information is available from multi-wavelength collaborations \citep{acc08,acc10}.  M87's core is variable, and optical and x-ray bands show common changes of about a factor of two, on timescales of months \citep{per03,har09}.  Very high energy (VHE) observations of variability on timescales of days \citep{aha06} suggest very compact emission regions on the order of the size of the inner accretion disk \citep{ner07}, and concurrent VHE, radio, and x-ray campaigns have helped tie the gamma ray emission from M87, for specific events, to areas close to the core \citep{abd09}.

	A number of models have been proposed in recent decades to describe accretion disks in AGN systems with low luminosity, compared to their Eddington luminosity (the luminosity limit at which the radiative pressure on the accreting matter balances the gravitational pull by the center body) -- an idea which has been referred to as a radiatively inefficient accretion flow (RIAF) \citep{yua03}.  Popular among these is the advection dominated accretion flow (ADAF) model, which cites the idea that, close to the horizon, most of the gravitational energy gained by particles is unable to radiate prior to them being advected onto the black hole \citep{nar94}.  The luminosity of M87 is about $L \approx 10^{-6} L_{Edd}$, where Edd represents the Eddington luminosity.  \citet{dim03} suggests an upper limit to the accretion rate of M87 around $\dot m = \dot M/\dot M_{Edd} = 1.6\times10^{-3}$, the Bondi accretion rate, based on gas properties derived from the Compton spectrum, where $\dot M_{Edd}$ is the accretion rate at which the Eddington luminosity is reached, assuming a radiative efficiency $\eta = 0.1$, representing the fraction of energy radiated by a typical particle of its total energy.  This would suggest that the efficiency of the source is $\eta \approx 10^{-5}$ if it accretes at $\dot M_{Edd}$, much lower than the canonical value $\eta = 0.1$ in a standard, efficient thin disk, making it a truly radiatively inefficient source.  A more recent estimate by \citet{lev11} based on calculated jet power and the capability of the system to extract power from a Kerr black hole suggests an accretion rate $\dot m = 10^{-4}$, for a maximally rotating black hole ($a/M = 1$).  Smaller spin values would then suggest higher accretion rates, scaling as $\dot m$ proportional to $a^{-2}$.  These rates may or may not be calculated at the black hole horizon, as some models choose other radii.  These accretion rate estimates can help set particle densities in radiative models.

	Models based on specific radiative mechanisms have been invoked recently to describe M87's spectrum.  For instance, \citet{ner07} suggest a scheme where electrons are accelerated by vacuum gap electric fields, in the black hole magnetosphere, while another paper makes use of centrifugal acceleration to heat electrons which upscatter ADAF disk photons \citep{rie08}.  Models such as these which propose novel acceleration methods are generally seeking to explain the heating of electrons to very high energies, which then, through the inverse Compton process, upscatter synchrotron photons to complete the x-ray and VHE spectrum \citep{har10}.
	
	Recent modeling work has attempted to restrict possible spin rates for M87 via a number of methods.  These papers typically use the rapid TeV variability to probe the black hole angular momentum, as in \citet{wan08}, where the TeV optical depth, assuming ADAF, is shown to strongly depend on the spin, and constrains it to greater than $a/M = 0.65$.  Similarly, the same group, in \citet{li09}, solve the relativistic hydrodynamical equations in the RIAF scheme to constrain the spin to greater than $a/M = 0.8$.  Advanced TeV imaging is likely to provide a very useful tool to tie down black hole spins in the near future.  For this paper, the lower limit on spin for M87 is assumed to be $a/M = 0.65$, to evaluate how well different spin rates fit the observed SED.
	
	We consider a scenario in which the immediate surroundings of the central black hole are responsible for the radio, infrared, and x-ray emissions observed.  This is due to emitting electron populations within the accretion flow and any GRMHD-driven outflows which HARM develops consistently.  Moreover, this region may prove to be the origin of VHE emission, though due to the very high energy electrons necessary to produce these through inverse Compton scattering, and the very low photon counts, this is a very difficult part of the spectrum to simulate via Monte Carlo (MC) methods.  Unlike other recent models focused on the radio-IR emissions, in this paper, we do not attempt to explain the energization method of electrons. Rather, we assume electron temperatures to be a free parameter proportional to the ion temperature due to the compressional heating inherent in MHD accretion methods.  The focus for this paper is on the dynamics specific to spin and accretion rates which produce appropriate Compton spectra.  This constant electron to ion temperature ratio is a common assumption \citep{gol05,mos09}, as there is no consensus on particle heating, and only work which is specifically related to heating mechanisms, such as \citet{shc10}, shows evidence against this.
	
	To motivate this constant temperature ratio assumption, recent PIC simulations \citep{zen05,zen07,lia09,liu11} demonstrated that magnetic reconnection and current sheet dissipation, which are believed to be the dominant kinetic processes dissipating MRI-driven turbulence, efficiently convert magnetic energy into hot electron thermal energy even in the absence of collisions.  Since the saturated MRI magnetic pressure given by MHD simulations scales with ion pressure, it is reasonable to expect the electron pressure heated by collisionless processes to scale with ion pressure.  Hence, as discussed above, for our models the electron temperature scales with ion temperature.
	
	This paper will focus on fitting data in the radio, IR, optical, and x-ray regimes (Section 2), particularly the \emph{Chandra}-band x-ray variability, of the SED by calculating physical properties via a general relativistic magnetohydrodynamic (GRMHD) accretion disk evolution scheme (Section 3.1) and applying output to a specialized MC radiation transport code (Section 3.2) by our specific modeling method (Section 4).  Discussion (Section 5) and Summary and Conclusions (Section 6) will contain interpretations of the data presented, detailing ramifications of results for clarifying the picture of M87's nucleus.  Suggestions regarding likely spin rate, accretion rate, and the mechanism of flaring will be presented, and reasonable modeling tasks for the future, based on the results, will be discussed.  Finally, an Appendix details modifications to the MC code for these types of sources, with highly anisotropic magnetic and velocity fields.

\section{Observations}

\begin{figure}
\includegraphics[scale=.17]{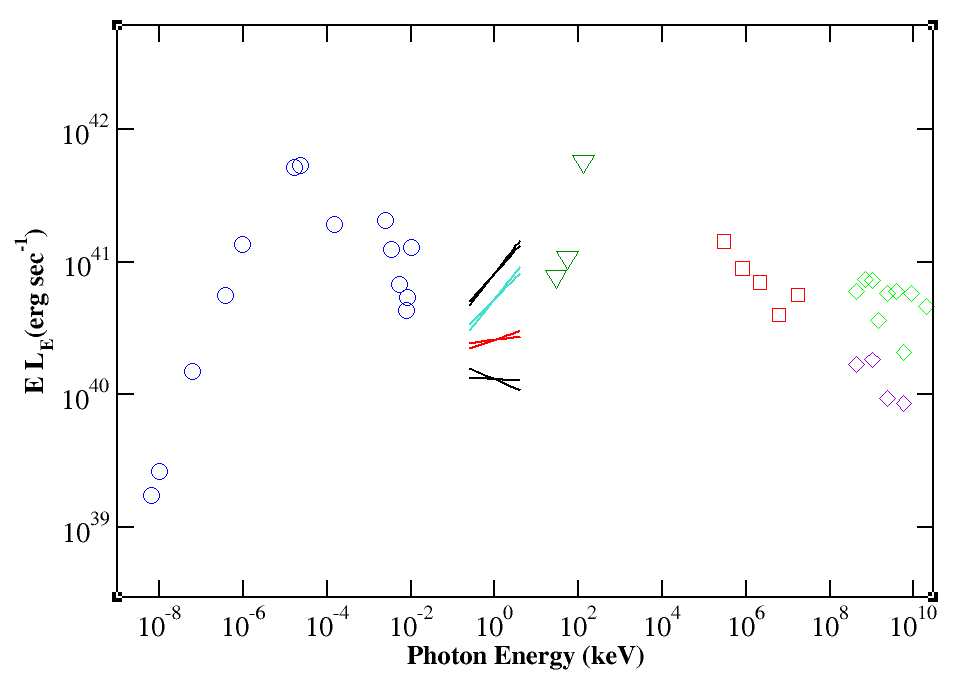}
\caption{SED of M87 in EL$_E$, to depict the variety of \emph{Chandra} x-ray indices.  Inverted triangles are \emph{Swift} long-term monitoring upper limits.  They will only be used to restrict fits to the average \emph{Chandra} spectrum (shown as the red bowtie), as flares are short-term transients.}
\end{figure}

\begin{deluxetable}{ccrrrrrrrrcrl}
\tabletypesize{\scriptsize}
\tablecaption{Table of \emph{Chandra} X-ray Spectra\label{tbl-1}}
\tablewidth{0pt}
\tablehead{
\colhead{Label} & \colhead{Date} & \colhead{Flux ($10^{40} erg/s$)} & \colhead{Spectral Index$^{a}$}
}
\startdata
Flare1 & 2008-2-16 & 8.24$\pm$0.13 & 0.62$\pm$0.031 \\
Flare2 & 2008-6-24 & 5.29$\pm$0.11 & 0.64$\pm$0.035 \\
Average & - & 2.59$\pm$0.055 & 0.92$\pm$0.044 \\
Quiescent & 2007-7-31 & 1.31$\pm$0.047 & 1.08$\pm$0.062 \\
\enddata
\tablecomments{\emph{Chandra} data is taken from the 0.2 to 6 keV band.}
\tablenotetext{a}{Index $\alpha$ for a power-law fit: $F_{\nu} \propto \nu^{-\alpha}$}
\end{deluxetable}

	M87 has been extensively observed throughout its energy range for decades.  Collected here is a full spectrum of data to describe its emissions, all plotted in Figure 1.  In the radio regime, data is available from the \emph{IRAM} Plateau de Bure interferometer \citep{des96} and the \emph{NRAO/VLA} \citep{bir91}.  At slightly higher energies, in IR, data is shown from \emph{Gemini Observatory}/OSCIR \citep{per01}, the \emph{Subaru Observatory}/COMICS, and \emph{Spitzer Space Telescope}/IRS/MIPS/IRAC \citep{per07}.  Next, in optical, \citet{bir91} presented data from the \emph{Palomar} telescope. These lower energy data are all represented as open circles in Figure 1.
	
	In hard x-ray, \emph{Swift/BAT} has provided upper limits from observations from 2005-2009 \citep{aje08,aje09} which are shown as inverted triangles in Figure 1.  Observations in VHE have also been collected, by HEGRA \citep{aha03,aha04}, H.E.S.S. \citep{aha06}, and \emph{Fermi}-LAT \citep{abd09}, with flaring behavior shown from H.E.S.S.  LAT data is shown as squares while H.E.S.S. flaring and quiescent data are depicted as diamonds.

	The most important data collected are from the \emph{Chandra} x-ray telescope, which are shown in Figure 1 as bowties.  Results were first given by \citet{wil02}, and variability data and descriptions of the observations and data used here is presented in \citet{har09}.  Private communications with Dan Harris, Francesco Massaro, and their group yielded spectral details which allowed for consideration of a variety of quiescent and flaring x-ray spectra.  Shown in Figure 1 and Table 1 are two flaring \emph{Chandra} x-ray spectra, a quiescent spectrum, and an average spectrum obtained by averaging the flux and power law spectral index ($\alpha$ for $F_{\nu} \propto \nu^{-\alpha}$) of all \emph{Chandra} data.  The highest flaring spectrum is significantly greater in flux than any other data point, so this paper will focus on fitting the second flaring point, which is more in line with the general trend of data.  So, any mention of the flaring \emph{Chandra} x-ray spectrum from this point on will refer to the second highest flaring point (Flare2 in Table 1).  The Flare1 bowtie will be left out of any further figures.

\section{Simulation Tools}

\subsection{HARM GRMHD Code}

\begin{figure}
\includegraphics[scale=.4]{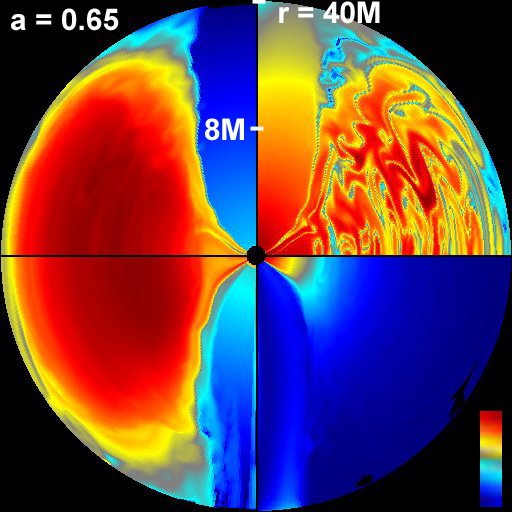}
\caption{Composite image of HARM output, showing data from runs with black hole spin a/M = 0.65, at time t = 2000M.  The top and bottom on the left show density and internal energy (temperature $\times$ density) plots, respectively.  Top and bottom on the right are magnetic field squared and bulk Lorentz factor, respectively.  Dark red corresponds to the highest normalized value for each, dark blue to the lowest.  Included are marks to depict the radial logarithmic spacing.}
\end{figure}

\begin{figure}
\includegraphics[scale=.4]{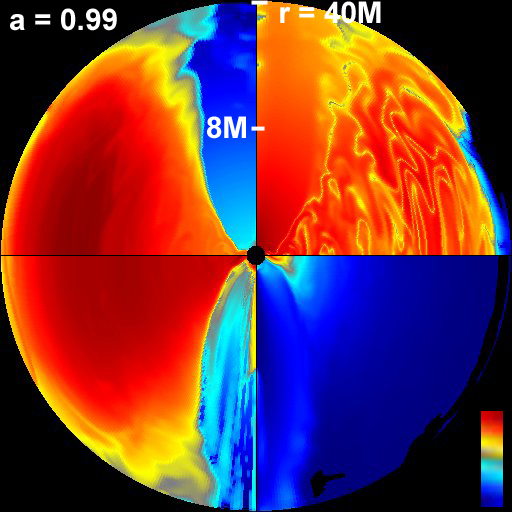}
\caption{Composite image of HARM output, as above, for a run of a/M = 0.99, at time t = 2000M.  Important to note is the dramatically stronger polar outflows, particularly visible in the internal energy (lower left) plot.  Only the higher spin runs show considerable emission contributions from this region.}
\end{figure}

	The physical values of the accretion disk system are calculated with the 2D axisymmetric HARM GRMHD code, which evolves an accreting black hole system based on a number of simple user-adjustable parameters as described in \citet{gam03} and \citet{nob06}.  From an initial torus perturbed from equilibrium by a small poloidal magnetic field, HARM integrates the GRMHD equations in a conservative scheme to consistently calculate parameters of the accretion flow.  Conserved variables are tracked by evaluating fluxes between simulation cells, and, from these, primitive physical variables such as particle density, internal energy, magnetic field, and velocity, are calculated.  For a full description of HARM'S algorithms and method, please see the cited papers \citep{gam03,nob06}.

	For our purposes, HARM is set-up with a small number of user-defined parameters, including adiabatic index, black hole spin value, simulation box size, torus position, and a small poloidal magnetic field to seed the torus.  From these initial parameters, the accretion disk evolves, governed largely by the magnetorotational instability (MRI), which describes the outward transport of angular momentum in the disk, and generates turbulence in the magnetic field from an initially poloidal field.  The physical space is divided into a spherical radial/angular grid, with cells spaced logarithmically in radius, and concentrated equatorially in the angular dimension.  This gives the highest resolution along the equator, at the horizon, where the shortest length scales of importance are located.
	
	In order to construct a useful library of LLAGN results from HARM, we have made a number of overlapping runs, all with an adiabatic index of 5/3, on grids of resolution 256x256 and 512x512.  These runs span a range of black hole spin value a/M = 0.65, 0.8, 0.9, and 0.99, the last being a near-maximally rotating black hole.  To check the effect of including larger simulation volumes, we have also made runs whose outer radii (in $GM/c^2$) range from 40 to 200.  Multiple simulation volumes can also be used to test for convergence of results given by the radiation transport code, given different volumes enclosed.  Emissivity curves are shown in Figure 4, to give an idea of the location of peak emission for different radiative mechanisms.  Important to note is that the bremsstrahlung curve peaks within r = 22M, whether larger volumes are considered or not, so most of the region's emissions will be reasonably modeled by using the smaller, better resolved, volume.
	
\begin{figure}
\includegraphics[scale=.17]{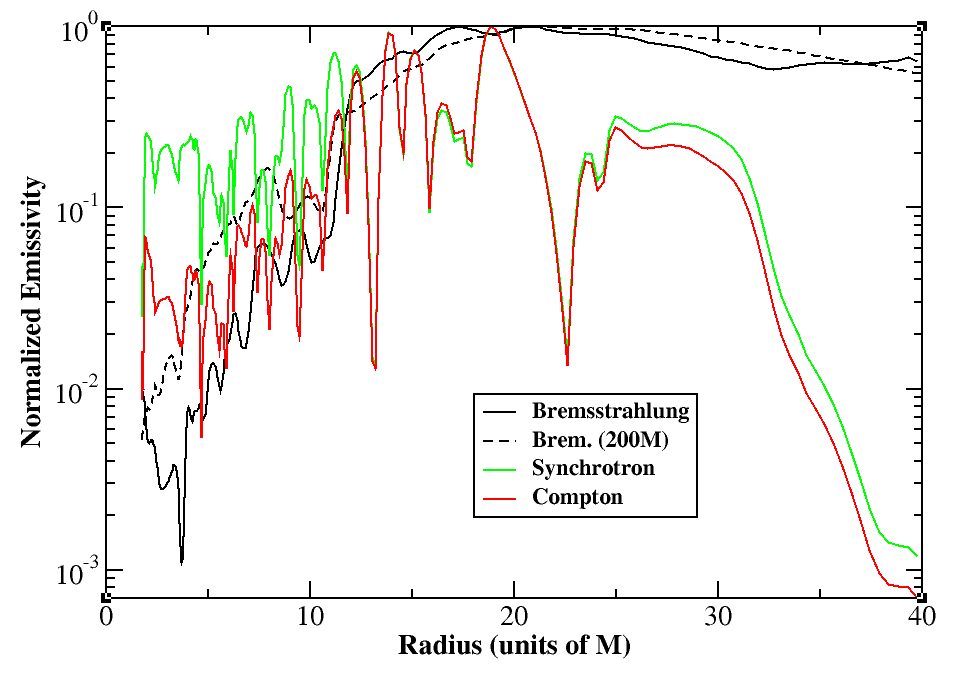}
\caption{Normalized emissivities for a typical a/M = 0.65 run.  Shown for comparison as the dashed line is a bremsstrahlung emissivity curve from a HARM run encompassing a volume out to r = 200M.  The curve labelled "Compton" is the generalized Compton emissivity, the synchrotron emissivity multiplied by the electron density.}
\end{figure}
	
	A brief note on the appropriateness of using 2D GRMHD for the problems being investigated: we contend that for our purposes of creating broadband spectra and constraining global parameters, the details of azimuthal modes would be averaged out even if included in full 3D, due to the rapid disk rotation in most of the relevant emission region.  That is, the global spectra of a 2D trial should look approximately the same as a 3D trial, given matching parameters.  This was noted by \citet{ohs05} in regard to Sgr A*, who stated that they checked that final MC-generated spectra were not significantly changed by averaging 3D MHD parameters over azimuth, implying that 3D effects may not be vital to conduct global spectral studies.  A primary difference in 3D and axially symmetric simulations is that MRI turbulence decays due to Cowling's anti-dynamo theorem throughout axially symmetric simulations.  Due to this, care is taken to select data at t = 2000 M (in black hole units) during the optimally turbulent time following initial in-fall, before the decay phase of the 2D turbulence. 

\subsection{Monte Carlo Radiation Transport Code}

The emission spectra based on physical parameters from HARM simulations are calculated by feeding the GRMHD data into our 2D axisymmetric Monte Carlo relativistic radiation transport code \citep{can87,lia88,boe01,fin07,che11}.  This simulation scheme allows bremsstrahlung and synchrotron emission, based on the radiative weight of each zone.  Emissions are then tracked through the simulation volume, with their energies and photon weights adjusted by absorption and scattering.

	All MC runs presented are on a 95x95 cell cylindrical grid, evenly spaced radially and vertically, in constrast to the spherical grid used by HARM (Figure 5).  The 95x95 grid is much finer than, for instance, 50x50 MC runs which present very similar results.  Based on a number of different mesh trials, the data are convergent at this scale.  The mapping procedure for physical values, from the HARM grid to the MC code grid, relies on averaging the values for all HARM cells that lie within each (usually much larger) MC code cell.  The number of MC photons (each representing a huge number of actual photons, reflecting the actual emission level of the zone) used for each run is 1 to 10 million.  Runs which needed more Compton scattering statistics relied on the photon splitting technique developed by \citet{che11}.  This significantly increases the quality of scattering statistics, allowing for both more consistent and shorter runs.
	
	This code has the capability to evolve electron distributions based on the Fokker-Planck (FP) equation.  Given that the electron-heating mechanism in LLAGN accretion disks is poorly understood and most likely due to collisionless plasma processes, we feel it is inappropriate to use the FP equation, so it is turned off for all trials.  As a first estimate, electrons are assumed to be thermal at a set temperature proportional to ions.  Future work will use particle-in-cell simulation results on the nonthermal heating of electrons by magnetic turbulence \citep{lia09,liu11}.
	
	In order to better model these types of sources, with highly anisotropic magnetic and velocity fields, modifications to the emission and scattering methods of the code were necessary and are detailed in the Appendix.

\begin{figure}
\includegraphics[scale=.45]{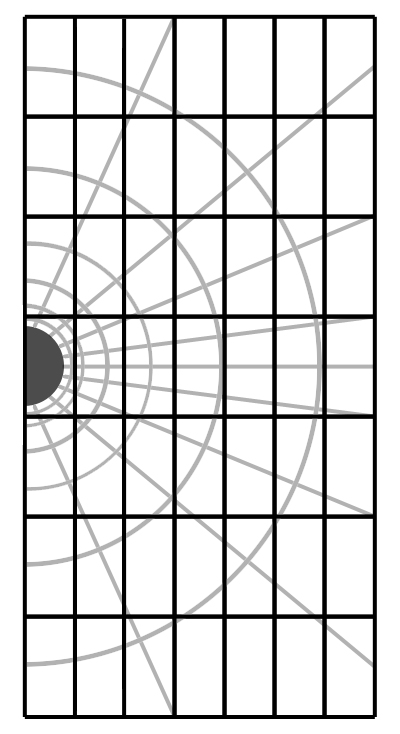}
\caption{Schematic depiction of the MC code (cylindrical, axisymmetric) grid overlaid on the HARM (spherical, axisymmetric, logarithmic) grid.  The HARM grid is much finer than the MC grid close to the horizon (shown as a solid semi-circle), and somewhat larger at large radii.  In this image, the respective grids are at appropriate ratios to one another through the simulation volume, though the horizon is exaggerated compared to the grid size.  In actual simulations, about seven MC cell lengths fit inside a Schwarzschild radius.  To model the horizon in the MC code, any cells within its radius are purely absorbing.}
\end{figure}

\section{Modeling}

	Evaluating HARM output to supply input data to the radiation transport code requires several steps, as described in \citet{hil10}:
	
1) All HARM units scale with a specified black hole mass, so the same runs may be applied to various astrophysical sources.  Specifying the black hole mass and a maximum density for the accretion flow yields values throughout the grid for MRI-saturated magnetic field components, ion temperatures due to adiabatic compressional heating, particle densities, and velocity components.

2) The MRI-saturated magnetic field values output by HARM are considered lower limits, as they do not include additional primordial fields (largely azimuthal) that may have been present in the plasma before its accretion.  Despite starting with a purely poloidal field, the azimuthal component of the field dominates due to the MRI evolution.  When scaling the magnetic field values for MC input, the amplitude is increased and components retain their respective ratios.  Because the azimuthal component is dominant to begin with, this approximation is equivalent to adding a primordial azimuthal field.

3) As the electron-heating mechanism in LLAGN accretion disks is poorly understood, a parametrized globally uniform electron-to-ion temperature ratio is applied, as in \citet{gol05} and \citet{mos09}.  This ratio is ultimately determined by collisionless (anomalous) heating processes, more efficient than Coulomb collisions.  This is acknowledged as a first approximation, and implies that the level of compressional heating of ions is proportional to the total heating of electrons, likely largely from magnetic dissipation.  In the future, for more advanced models, we will add a small nonthermal (power-law) component to the thermal population to model the VHE data.

	So, given a black hole mass, maximum density, magnetic field value (over the MRI-saturation value), and electron-to-ion temperature ratio, HARM output can be used to compute radiation output.  In this case, the maximum value of each parameter is set, and each other cell's value scales accordingly.

\subsection{Spectral Modeling Results}

	Typical HARM data was taken at $t \approx 2000 M$, before accretion-driven turbulence dies down.  At this point, we used the HARM output as input to the MC spectral modeling.  The particle densities chosen for models are based on accretion rates suggested in literature. When the maximum particle density is $n = 1\times10^7$ cm$^{-3}$, the maximum accretion rate within the simulation volume is $\dot{m}_{max} \approx 10^{-4}$.  This is the case for all spin rates, while the accretion rate through the horizon ranges from $\dot{m}_{H} \approx 2\times10^{-6}$ up to $\dot{m}_{H} \approx 2\times10^{-5}$, depending on the specific model -- higher spin rates have correspondingly lower horizon accretion rates, due to outflows.  As the maximum matches the accretion estimates of recent work, this was chosen as a benchmark for our models.  For the rest of the paper, accretion rates will be given as maximum values as these are similar between models with differing spin rates.  In order to evaluate the impact of a higher or lower accretion rate, two other maximum densities were chosen:  $n = 3\times10^6$ cm$^{-3}$ and $n = 3\times10^7$ cm$^{-3}$, for maximum accretion rates of $\dot{m}_{max} \approx 3\times10^{-5}$ and $\dot{m}_{max} \approx 3\times10^{-4}$, respectively.  Full trials were then performed using these three densities, leaving two parameters for adjustment: electron temperature and magnetic field.

	As the main interest in fitting spectra is to evaluate the origin of flaring mechanisms, runs are chosen for their fits to x-ray data.  The starting point for each density is then to fit the average x-ray spectrum, whose flux and index are averaged over all \emph{Chandra} x-ray data, not including those with possible pile-up.
	  
	Figure 6 below shows the effect of changing each parameter (density, temperature, and magnetic field) by a factor of two.  Obviously, temperature and density have a significant impact on the spectral shape at the x-ray spectrum, each hardening the spectrum when raised.  Conversely, the magnetic field, in general, uniformly changes the flux throughout the x-ray spectrum, without changing the x-ray spectral index considerably.  This is because increasing the magnetic field increases the flux of the synchrotron curve at the same rate it increases that of the Compton components, as the upscattered photons are synchrotron in origin.
	
\begin{figure}
\includegraphics[scale=.17]{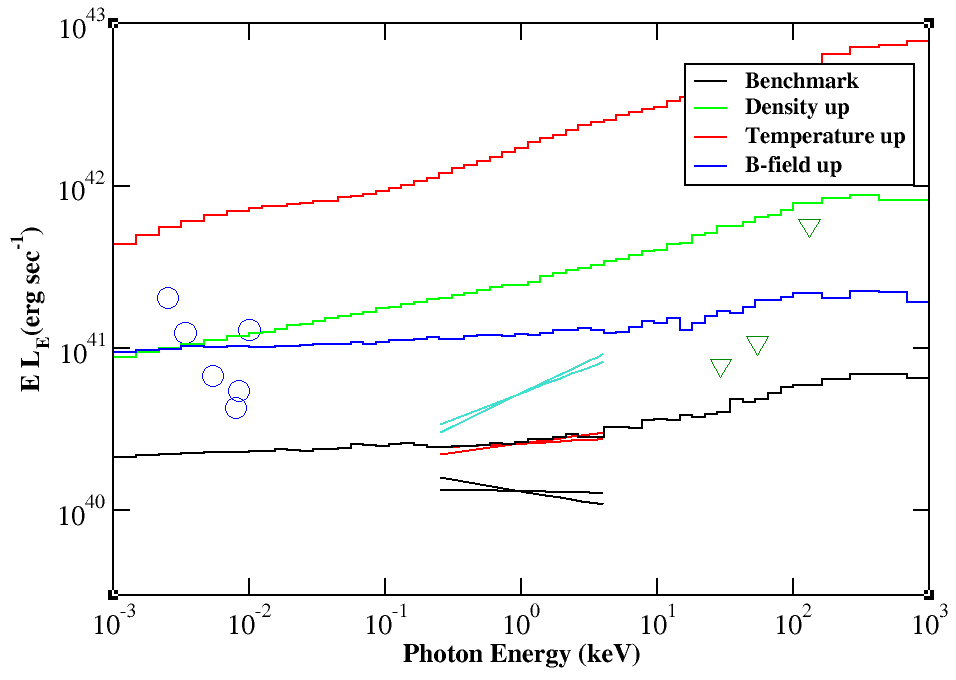}
\caption{A benchmark fit at a/M = 0.9, $n = 1\times10^7$ cm$^{-3}$, and three spectra generated by alternately raising a single parameter by a factor of two.  This is shown in EL$_E$, to better depict index changes in x-ray.}
\end{figure}
	
	As the density in each trial is fixed, this means that the obvious method of fitting spectra is to vary temperature to fit spectral index while varying magnetic field to fit flux, until the average x-ray bow-tie is satisfactorily fit.  It should be noted that, in general, higher black hole spin rates lead to higher densities at small radii, where velocities of the accreting matter are much greater.  Higher velocities lead to harder spectra due to Doppler boosting, so higher spin trials have lower indices, for similar parameters.
	
	Once full trials were completed to fit the average x-ray spectrum -- for each of the three density points, and for each of the four black hole spins -- the quiescent and flaring spectra needed consideration.  Given that the flaring mechanism is unknown, the simplest changes to interpret involve varying a single parameter each time.  Specifically, if a change in accretion rate is responsible for the flaring behavior, we approximate it by a global density change at fixed temperature.  If an increase in electron heating is responsible, we model this by a global temperature change at fixed density.  For this reason, a full suite of trials has been done which fit the quiescent and flaring x-ray spectra by changing merely one of these (maximum density, maximum electron temperature) from the benchmark model which fit the average x-ray spectrum.

	Since the $n = 1\times10^7$ cm$^{-3}$ runs have the closest accretion rate to that suggested in literature  ($\dot{m}_{max} = 10^{-4}$) these runs were evaluated first.  The fits for each spin rate are normalized to match the flux of the average x-ray spectrum.  In general, these are poor fits to radio, IR, and optical data.  Because the higher spin rates lead to larger densities at higher accretion velocities, and therefore harder spectra, the a/M = 0.99 trial had to use the lowest temperature value, and, conversely, the 0.65 trial the highest, to fit the slope of the x-ray data.  This leads to the lower spin rates providing better fits at low energies, as the synchrotron flux is higher.  However, none of these adequately fit any of the low energy spectrum, so the quiescent and flaring fits are not considered.

	The fits for the $n = 3\times10^6$ cm$^{-3}$ runs, which yields an accretion rate lower than suggested by literature, are qualitatively similar to those discussed above.  They fall short at the radio-IR range; therefore, quiescent and flaring trials are again not considered.

\subsection{Fits using a density of $3\times10^7$ cm$^{-3}$}

\begin{figure}
\includegraphics[scale=.17]{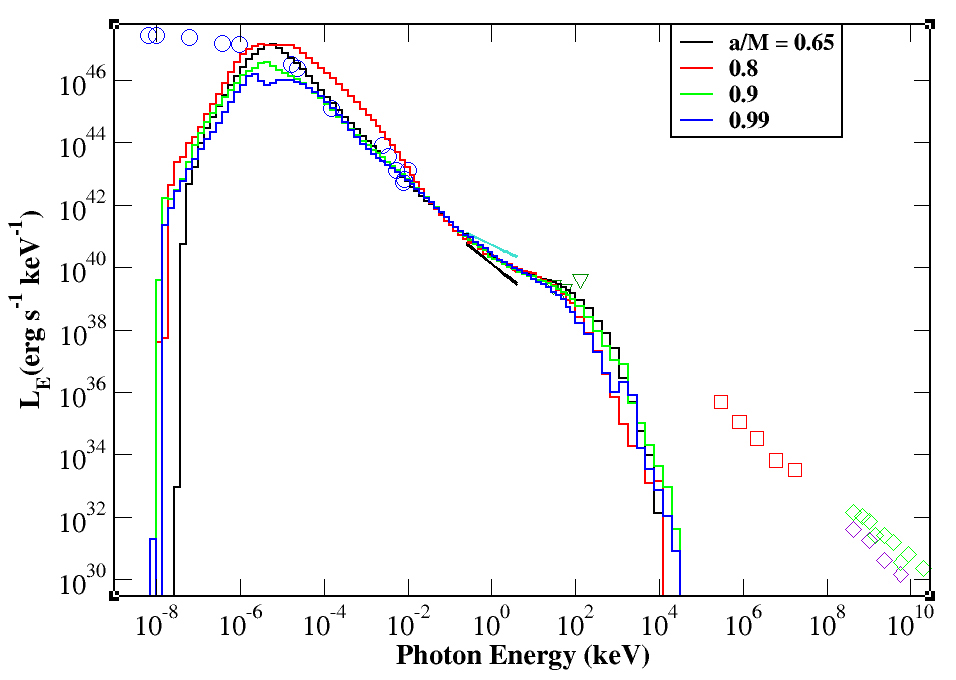}
\caption{$n = 3\times10^7$ cm$^{-3}$ runs.  Shown below are quiescent and flaring fits, as these runs describe lower energy data best.}
\end{figure}

	The third set of fits uses a density of $n = 3\times10^7$ cm$^{-3}$.  This corresponds to an accretion rate above the recently quoted value, but still well below the Bondi accretion rate which has been suggested as an upper limit to the level of accretion.  The Bondi accretion rate defines spherical accretion onto a compact object, $\dot M = \pi R^2 \rho v$, where $\rho$ and $v$ are the density and sound speed, respectively, of accreting matter, and $R$ is the characteristic radius found by equating the object's escape velocity and relevant sound speed \citep{dim03}.
	
	As seen in Figure 7, these spectra offer good fits to radio, IR, and optical data, unlike lower density trials.  The second big change is the visibility of the shoulder of the bremsstrahlung emission in hard x-ray, above \emph{Chandra} energies.  This is the first time here the upper limits from Swift data \citep{aje08,aje09} require consideration.  As Swift data are averaged limits over several years, they only restrict fits to the average \emph{Chandra} x-ray spectrum, but they are still right at the edge of all the spectra with this density.  Essentially, as bremsstrahlung emission scales as density squared, this puts a limit on maximum model density at $n = 3\times10^7$ cm$^{-3}$.

	While the a/M = 0.8 trial overestimates much of the low energy data, three of the four fits shown above are approximately equally good through the radio, IR, and optical bands.  All also fit with a nearly pure power-law at the average x-ray spectrum, and come close to the Swift x-ray upper limits.  The only fit which lies comfortably beneath the Swift upper limits is the a/M = 0.99 run.  This is due again to the fact that with higher spin runs, the emitting/scattering electron populations are moving with higher maximum bulk velocities.  This means that the maximum temperature can be turned down considerably while still maintaining the appropriate x-ray index, thereby moving the bremsstrahlung cut-off to a substantially lower energy than the other trials.
	
	The quiescent and flaring fits shown in Figure 8 are for the runs with a/M = 0.9.  Shown as (a) is the average fit shown in the zoomed out image above (Figure 7).  As detailed in the fitting methodology, the quiescent and flaring fits are changed from the average data fit in only one parameter: either temperature or density.  Spectra (c) and (e) are changed only in temperature from the benchmark (a), while (b) and (d) are changed only in density.
	
\begin{figure}
\includegraphics[scale=.17]{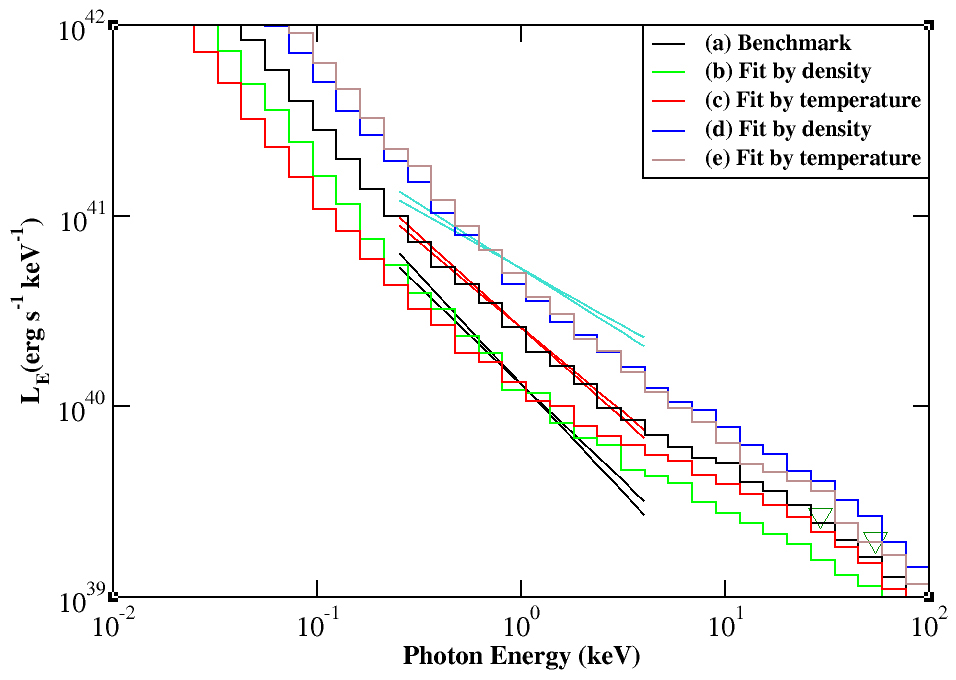}
\caption{a/M = 0.9, with $n = 3\times10^7$ cm$^{-3}$.  These views are zoomed in to focus on the x-ray spectra, to better show changes in index here.  (a) is the average fit shown above.  Fits to quiescent data, (b) and (c) are varied from (a) in density and temperature, respectively.  Flaring fits (d) and (e) are varied in density and temperature, respectively.  That is, (c) and (e) are changed only in temperature from (a), and similarly for (b) and (d) in density.}
\end{figure}	
	
	Either quiescent trial could be seen to fit the quiescent spectrum reasonably; its index is quite similar to the average spectrum.  The bremsstrahlung shoulder is more visible at this energy than for the average trial, and because the density fit drops this a bit lower than the temperature fit, it maintains the quiescent slope better.
	
	The flaring trials are, at first glance, quite poor.  The amount of change in index to the flaring spectrum is much more noticeable than to the quiescent spectrum.  Again, the bremsstrahlung cut-off plays a large role in these fits.  Turning up the temperature does not get the bremsstrahlung slope up to the flaring spectrum, but turning up the density does.  Because of this, it can be seen that a small change in density can yield a large change in x-ray spectral index.  Specifically, the lower energy spectrum for the flaring x-ray density fit has a much higher index, while the higher energy portion has an index quite close to the flaring spectrum.  
	
	In general, the fits by changing density are better at fitting the quiescent and flaring data spectra, largely due to the fact that the extreme index change to the flaring spectrum can be explained by the presence of the bremsstrahlung bump.  The other spin values considered, a/M = 0.65, 0.8, and 0.99, yielded similar results to the previous trials.  Most noticeable in each case is that the density fits are significantly better than the temperature fits, suggesting again that accretion rate variations may be more reasonable to suggest as the dominant flaring mechanism.	
	
\section{Discussion}

	The sample fits immediately suggest that a density higher than $n = 1\times10^7$ cm$^{-3}$ is necessary to yield an adequate fit to the radio, IR, and optical spectra.  The $n = 3\times10^7$ cm$^{-3}$ runs have some conflict with the Swift upper limits in the hard x-ray regime, which restricts the accretion rate to $\dot{m}_{max} = 3\times10^{-4}$.  These upper limits are only considered for the average \emph{Chandra} x-ray data fits, as they are essentially averages over a number of years.
	
	As it is expected that a larger simulation volume for the accretion flow would only add significantly to the bremsstrahlung flux, $n = 3\times10^7$ cm$^{-3}$ can be seen as an upper limit to the density maximum.  This is because of the artificial initial condition of a small-radius torus, rather than near Bondi-scale accretion.
	
\begin{deluxetable}{ccrrrrrrrrcrl}
\tabletypesize{\scriptsize}
\tablecaption{Table of Model Average X-ray Fits for $n = 3\times10^7$ cm$^{-3}$\label{tbl-1}}
\tablewidth{0pt}
\tablehead{
\colhead{Label} & \colhead{Spin (a/M)} & \colhead{Magnetic Field} & \colhead{Field Scaling$^{a}$} & \colhead{Temperature} & \colhead{Density} & \colhead{Fit to:} & \colhead{\emph{Chandra} Index$^{b}$} & \colhead{Model Index$^{b}$}
}
\startdata
1 & 0.65 & 2000 G & 62.5 & 15 MeV & 3$\times10^7$ cm$^{-3}$ & Average & 0.92$\pm$0.044 & 0.90\\
2 & 0.8 & 2600 G & 47.3 & 20 MeV & 3$\times10^7$ cm$^{-3}$ &  Average & 0.92$\pm$0.044 & 0.85\\
3 & 0.9 & 500 G & 6.1 & 22 MeV & 3$\times10^7$ cm$^{-3}$ &  Average & 0.92$\pm$0.044 & 0.92\\
4 & 0.99 & 460 G & 2.9 & 7 MeV & 3$\times10^7$ cm$^{-3}$ &  Average & 0.92$\pm$0.044 & 0.86\\
\enddata
\tablecomments{The indices shown are in the energy band from 0.2 to 6 keV, both for \emph{Chandra} data and model fits.  Magnetic field, electron temperature, and electron density values given are the maximum for each within the simulation grid, which all other cells scale to.}
\tablenotetext{a}{This value is the factor the GRMHD MRI-saturated magnetic field had to be scaled by to appropriately normalize MC output, as discussed in the text.}
\tablenotetext{b}{Index $\alpha$ for a power-law fit: $F_{\nu} \propto \nu^{-\alpha}$}
\end{deluxetable}	
	
	The details of the average x-ray fit benchmark MC trials are shown in Table 2.  These include the physical parameters of the runs (electron temperature, density, and magnetic field), \emph{Chandra} spectrum fit to, and data and model spectral indices.
	
	As discussed above, we focus on the $n = 3\times10^7$ cm$^{-3}$ trials as these gave the best fits to radio-IR-optical data.  There are four different spin rates to consider: a/M = 0.65, the lower limit suggested in literature, 0.8, 0.9, and 0.99, a near-maximally rotating black hole.  The fits to \emph{Chandra} spectra don't allow much differentiation between these trials, as they yield very similar results.  Spectral indices range from $\alpha$ = 0.85 to 0.92, close fits to the \emph{Chandra}-given 0.92.
	
	Also of interest in Table 2 are the specific parameters required for fits.  The general trend is that lower spin runs require higher electron temperatures and magnetic fields to match appropriate spectral properties.  HARM runs conducted to test the response of the simulation to additional primordial toroidal fields have shown the MRI development is approximately the same (with higher final field values) for field scaling up to an order of magnitude.  Beyond this, the large magnetic pressure dominates the simulation, inhibiting accretion.  This allows an easy evaluation of the average fit models, as the 0.65 and 0.8 spin runs require much higher field scaling, while the 0.9 and 0.99 trials are more reasonable.
	
\begin{deluxetable}{ccrrrrrrrrcrl}
\tabletypesize{\scriptsize}
\tablecaption{Table of Model Flaring/Quiescent X-ray Fits for $n = 3\times10^7$ cm$^{-3}$\label{tbl-1}}
\tablewidth{0pt}
\tablehead{
\colhead{Label} & \colhead{Spin (a/M)} & \colhead{Scaling factor} & \colhead{Fit to:} & \colhead{Fit by:} & \colhead{\emph{Chandra} Index$^{a}$} & \colhead{Model Index$^{a}$}
}
\startdata
5 & 0.65 & 0.87 & Quiescent & Temperature & 1.08$\pm$0.062 & 0.67\\
6 & 0.8 & 0.75 & Quiescent & Temperature & 1.08$\pm$0.062 & 0.55\\
7 & 0.9 & 0.86 & Quiescent & Temperature & 1.08$\pm$0.062 & 0.70\\
8 & 0.99 & 0.80 & Quiescent & Temperature & 1.08$\pm$0.062 & 0.64\\
\\
9 & 0.65 & 0.83 & Quiescent & Density & 1.08$\pm$0.062 & 0.80\\
10 & 0.8 & 0.77 & Quiescent & Density & 1.08$\pm$0.062 & 0.99\\
11 & 0.9 & 0.83 & Quiescent & Density & 1.08$\pm$0.062 & 0.88\\
12 & 0.99 & 0.77 & Quiescent & Density & 1.08$\pm$0.062 & 0.87\\
\\
13 & 0.65 & 1.13 & Flaring & Temperature & 0.64$\pm$0.035 & 0.99\\
14 & 0.8 & 1.25 & Flaring & Temperature & 0.64$\pm$0.035 & 1.00\\
15 & 0.9 & 1.14 & Flaring & Temperature & 0.64$\pm$0.035 & 1.02\\
16 & 0.99 & 1.19 & Flaring & Temperature & 0.64$\pm$0.035 & 1.03\\
\\
17 & 0.65 & 1.22 & Flaring & Density & 0.64$\pm$0.035 & 0.93\\
18 & 0.8 & 1.33 & Flaring & Density & 0.64$\pm$0.035 & 0.84\\
19 & 0.9 & 1.20 & Flaring & Density & 0.64$\pm$0.035 & 0.94\\
20 & 0.99 & 1.27 & Flaring & Density & 0.64$\pm$0.035 & 0.89\\
\enddata
\tablecomments{The scaling factor given is the factor either the electron temperature or electron density is multiplied by from the average benchmark fit (see Table 2) to get a new value for the quiescent or flaring fit presented.}
\tablenotetext{a}{Index $\alpha$ for a power-law fit: $F_{\nu} \propto \nu^{-\alpha}$}
\end{deluxetable}

\subsection{Flaring and Quiescent Fits}

	Table 3 details quiescent and flaring spectral fits, as well as describing the change necessary for each fit, from the benchmark average fits, for each spin value.  The 0.65 and 0.8 trials details are included for completeness, but will not be discussed extensively.  As mentioned previously, their magnetic field scaling values suggests they may not be reasonable spin values, and the qualitative analysis is very similar to that of higher spin rate trials.
	
	The quiescent spectrum is very close in spectral index to the average spectrum.  It can be fairly easily fit by decreasing density from the average fit's parameters.  Lowering temperature yields a slightly less satisfactory fit at the quiescent spectrum, with a spectral index too low, and spectrum too hard.  For every trial conducted, regardless of spin, the density fits showed closer fits to \emph{Chandra} data (with quiescent spectral index $\alpha$ = 1.08) than the temperature fits.  Specifically, the 0.99 spectrum (benchmark $\alpha$ = 0.86) became slightly softer ($\alpha$ = 0.87) as expected for the density adjustment, while scaling temperature led to a harder spectrum ($\alpha$ = 0.64).  The 0.9 spectrum (benchmark $\alpha$ = 0.92) became harder with density scaling ($\alpha$ = 0.88), but this is still significantly better than the temperature-adjusted trial ($\alpha$ = 0.70).
	
	Flaring fits are more complicated to achieve.  Because the spectral index is quite a bit lower than that of the average fit -- and the change between indices is much greater than between the average and quiescent -- it is nearly impossible to fit the flaring spectrum by simply adjusting one parameter.  However, the fact that the bremsstrahlung emission is visible here, whereas it wasn't in the lower density trials, means that density changes can have a large impact on where in the energy band the change from Compton spectrum to bremsstrahlung spectrum occurs.  At this density, $n = 3\times10^7$ cm$^{-3}$, the bremsstrahlung spectrum dominates at an energy around several keV.  Therefore, when the density is increased, and the bremsstrahlung component is increased more than the Compton component, the bremsstrahlung emission is visible down closer to 1 keV.  
	
	To demonstrate, the bremsstrahlung spectrum above 1 keV (for the run with $n = 3\times10^7$ cm$^{-3}$, a/M = 0.9, average fit) much better describes the hard index at the flaring spectrum than the softer Compton spectrum.  Specifically, a power law fit from 0.2 to 1 keV has an index $\alpha$ = 1.13, while a fit from 1 to 6 keV has an index $\alpha$ = 0.69.  Fitting  the full range from 0.2 to 6 keV yields an index $\alpha$ = 0.92.  These compare to \emph{Chandra} x-ray spectral indices of 1.08 for the quiescent spectrum, 0.64 for flaring, and 0.92 for average.  As the \emph{Chandra} x-ray data ranges from 0.2 to 6 keV, it is clear that small changes in parameters could lead to any of the three of these fits being appropriate throughout the range.
	
	None of the fits shown exactly traces the flaring spectrum, but it is simple to see that the density variations work better than temperature variations, and the density fits show promise at slightly higher energies to fitting the flare spectral index.  Quantitatively, the \emph{Chandra} flaring data (with spectral index $\alpha$ = 0.64) is better fit by density changes ($\alpha$ = 0.94, 0.89 for spin 0.9, 0.99) than by temperature changes ($\alpha$ = 1.02, 1.03 for spin 0.9, 0.99).  For both spin rates, density fits are more consistent with data.
	
\subsection{Model Tests and Evaluation}	
	
	With both flaring and quiescent spectra better fit from the average spectrum by density changes, it is worth considering how much the density has to be changed for these fits.  From the starting density of $n = 3\times10^7$ cm$^{-3}$, the quiescent spectrum was best fit with an average of $n = 2.4\times10^7$ cm$^{-3}$.  Similarly for the flaring spectrum, an average of $n = 3.7\times10^7$ cm$^{-3}$ was required.  This suggests changes in accretion rate, from the average fit, of about 20-25\%.  As discussed previously in \citet{hil10}, mass accretion rates vary in HARM trials by about a factor of two.  Similarly, \citet{dex09,dex10} suggest variability up to about 50\% for both 2D and 3D Sagittarius A* models.  Both of these examples comfortably allow for the density variations required for fits.
	
	Following \citet{mos11}, we consider the size of the 230 GHz photosphere from our models, to compare to VLBI measurements by \citet{fis10} which found structure at this frequency on the scale of several Schwarzschild radii.  All four spin trials had photospheres within 10M, with higher spin trials having smaller photospheres, as expected.  Clearly, accretion flow models are consistent with current VLBI results.
	
	The assumption that radiative cooling is unnecessary in the GRMHD calculation is motivated by the flow being radiatively inefficient.  The typical 0.9 spin average run has a radiative efficiency of $\eta \approx 10^{-2}$, an order of magnitude less than the canonical value $\eta \approx 10^{-1}$ for an efficient thin disk.
	
	Based on the trials done, there is little to choose between the different spin rates considered.  The 0.9 and 0.99 runs are more likely than lower spin trials due to the primordial magnetic fields required.  Of these, the 0.9 run may be marginally better at low energies, but not definitively so.  On the other hand, the density changes are definitely better than the temperature changes, suggesting that changes in accretion rate are most likely to explain flaring behavior, based on these trials.
	
\section{Summary and Conclusions}

	In order to explore flaring mechanisms at play in M87's core, full trials have been conducted using a GRMHD accretion evolution scheme, to solve for global physical parameters, and a novel MC radiation transport code, to generate spectra from these parameters.  The flaring data being displayed is in \emph{Chandra}'s x-ray band.  Trials are fit to an average x-ray spectrum, and then changes necessary to fit quiescent and flaring x-ray spectra are discussed, along with ramifications of specific changes.
	
	To evaluate likely spin rates, with literature suggesting a/M $\geq$ 0.65, four different GRMHD runs are used, with a/M = 0.65, 0.8, 0.9, and 0.99.  Articles also suggest an accretion rate under $\dot m = 1.6\times10^{-3}$, but above or around $\dot m = 1\times10^{-4}$.  To take this into account, the maximum density assigned was adjusted to provide sets of runs at $\dot{m}_{max} \approx 3\times10^{-5}, 1\times10^{-4}$, and $3\times10^{-4}$.
	
	Only the highest accretion rate trials, which correspond to a maximum density of $n = 3\times10^7$ cm$^{-3}$, manage to fit lower energy data adequately, and so were focused upon for fitting the flaring x-ray spectrum.  This density also shows that higher average accretion rates are unlikely, as the bremsstrahlung emission is very close to upper limits provided by the Swift hard x-ray data.  As including larger volumes can only maintain or raise the bremsstrahlung flux, this places an upper limit on maximum density and accretion rate.
	
	Quiescent and flaring fits were presented which require only changing density or temperature from the average fits.  This can simulate either a global accretion rate change, or a global electron temperature change -- indicative of more efficient electron heating.  During none of these trials was the magnetic field changed in fitting quiescent and flaring spectra, in order to isolate the parameter changes.
	
	The quiescent x-ray spectrum has a very similar spectral index to the average x-ray spectrum.  Because of this, it is fairly simple to get a close fit by simply dropping either temperature or density from the average spectrum.  In general, no spin rate stands out as having outstanding quiescent fits.  They all exhibit similar behavior: the density-changed trials have a slightly better index, while the temperature-changed trials are a bit too hard at the quiescent spectrum.
	
	The flaring spectrum is more difficult to explain.  Unlike the quiescent spectrum, the flaring spectrum's index is substantially harder than that of the average spectrum, suggesting a much harder spectrum.  Again, no spin rate displays perfect fits.  These actually look worse than the quiescent fits, because in order to explain both the slope and flux changes, the bremsstrahlung bump has to be enhanced.  This leads to a transition between Compton and bremsstrahlung dominance essentially right at the x-ray data, so that small changes can lead to the x-ray points falling on either side of this transition.  In general, the higher energy (bremsstrahlung) side of the x-ray runs seems to adequately describe the index at the flaring spectrum, while the lower energy (Compton bumps) side traces the average and quiescent spectra well, but this is very parameter-sensitive.
	
	Overall, the a/M = 0.9 and 0.99 spin runs are marginally better than lower spins at fitting all three x-ray spectra considered, and the $n = 3\times10^7$ cm$^{-3}$ trials were the only ones which provided a good fit to radio-IR data at all.  This suggests that a maximum accretion rate $\dot{m}_{max} \leq 3\times10^{-4}$ and spin of a/M $>$ 0.8, both well within limits established in literature, are the most appropriate for the core of M87.  Scaling between the average x-ray spectrum and flaring and quiescent spectra requires only simple changes in accretion rate ($\approx 20\%$).
	
	It may be important to note that the higher spin trials have the most prominent polar outflows and show significantly more emission from this assumed jet base than lower spin rates.  These trials are entirely thermal, though jet emission is likely to be non-thermal, and as \citet{dex11} show, the core spectrum can also be fit with non-thermal jet-dominated or jet-and-disk models.
	
	Future modeling work on this source will focus both on more detailed fits of the data already considered, as well as fits including Fermi and VHE data, which was not used in this paper.  Continued observations of M87 by \emph{Chandra} and \emph{Swift}, which can be used to confirm the trend seen in flux vs. index, can help to prove the validity of this work.  Specifically, observation of two clear trends can show that the bremsstrahlung and Compton components do both need to be included to fit the flaring spectrum depicted.
	
	Furthermore, more accurate and consistent electron heating mechanisms, involving particle-in-cell simulation results, should lead to better-described electron distributions.  Recent results have suggested that particle acceleration by magnetic reconnection in similar situations to MRI disks display a dual Maxwellian nature -- with one major population at a low energy, and a higher temperature second population.  As the spectral indices already considered should be fairly appropriate extended to the VHE regime, this seems quite promising to describe the spectrum more completely.  However, since the Fermi-VHE spectral index is softer than the \emph{Chandra} x-ray index, any additional nonthermal electron component invoked to model those high energy data will not impact the thermal spectral fitting to the lower energy data performed here.
	
\acknowledgments

GLH would like to specially thank Xuhui Chen, Dan Harris, and Francesco Massaro, for beneficial discussion and sharing of techniques and observations.

Both authors are also indebted to the anonymous referee, whose thoughtful comments and questions have made a significant impact on the content and quality of this work.


%% in parentheses, after the keyword, but they will not be verified.



\appendix

\section{Monte Carlo Code Modifications}

\begin{figure}
\includegraphics[scale=.33]{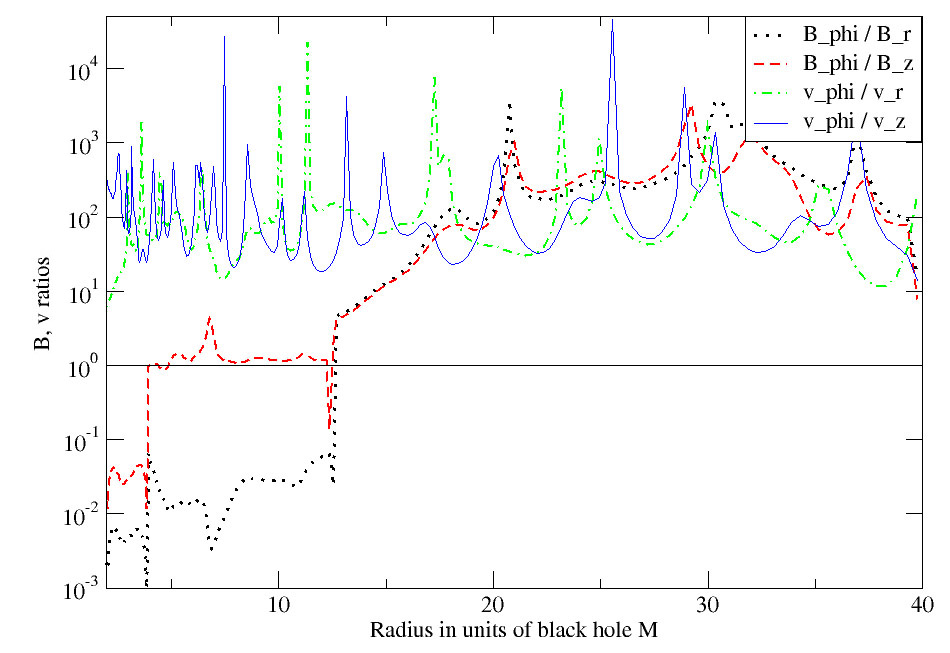}
\caption{Ratios of components of magnetic field and velocity for a HARM GRMHD run of black hole spin a/M = 0.99.  The line at unity emphasizes the high degree of anisotropy in these components.}
\end{figure}

	Data from HARM suggest two shortcomings to the MC code, namely the anisotropy in velocity and magnetic field.  As shown in Figure 9, for a typical run of a/M=0.99, the components of these parameters are usually very disparate.  This will obviously lead to highly anisotropic synchrotron radiation and scattering characteristics, for adequately high fields and velocities.  The MC code previously considered synchrotron emission as angle-independent and did not allow for relativistic beaming, boosting, and scattering.  In order to create a tool as consistent as possible for a number of astrophysical sources, these issues needed to be addressed in the MC code.

\subsection{Anisotropic Magnetic Field}

	For fields of the magnitude expected in AGN accretion disks, the dominant effect is on the direction of emission of synchrotron radiation.  By \citet{pet81} the emission scales as $e^{-\frac{\nu}{\nu_b}[\frac{4.5}{\sin{\theta}}(\frac{\nu_b}{\nu kT})^{2}]^{1/3}}$ for semi-relativistic temperatures and $\frac{1+\cos^{2}\theta}{\sin^{2}\theta}e^{-(\nu/\nu_b) ln(2\nu_b/e\nu kT\sin^{2}\theta)}$ for non-relativistic temperatures, where $\nu$ is the photon frequency, $\nu_b = \frac{eB}{2\pi m_e c}$ is the gyrofrequency, $B$ is the magnetic field, $T$ is the electron temperature, and $\theta$ is the angle between photon travel and field direction.  In an accretion disk, the toroidal magnetic field is often highly dominant.  The impact is that photons are emitted strongly perpendicular to this direction.
	
	To best depict the effect, Figure 10 shows a run with a nearly purely radial magnetic field.  This results in a much larger number of photons emitted perpendicular to rather than parallel to fields, which leads to a greater flux in the polar direction than the equatorial direction.

\begin{figure}
\includegraphics[scale=.23]{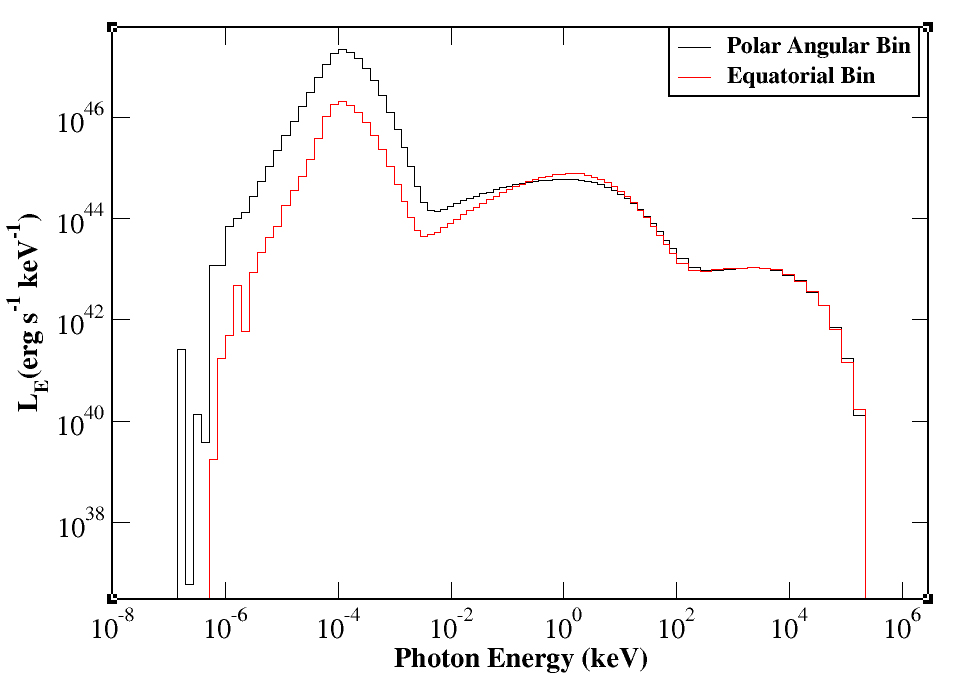}
\caption{Trial of the anisotropic magnetic field modifications.  This is a single zone run with nearly purely radial field, to best show the effect of the changes.  As expected, the synchrotron emission is strongly anisotropic, while the Compton components are much more isotropized.}
\end{figure}

\subsection{Anisotropic Velocity Field}

	As the impact will be seen in both emission and scattering events, the process of including plasma flow velocity requires changes to several of the MC code's routines:
1) All emission is beamed in the direction of relativistic plasma flow, and Doppler boosted, with $\cos{\phi^\prime} = \frac{\cos{\phi} - \frac{v}{c}}{1 - \frac{v}{c} \cos{\phi}}$ and $\nu^\prime = \frac{\nu(1 - \frac{v}{c} \cos{\phi})}{\sqrt{1 - (\frac{v}{c})^{2}}}$, where $\phi$ is the angle between the photon and bulk flow directions, $\nu$ is the photon frequency, $v$ is the bulk flow magnitude, and $\prime$ represents the bulk flow frame.

2) Compton scattering frequency increases when photons travel against the flow of particles (head-on), and decreases when moving with the flow (tail-on), as $f^\prime = f(1 - \frac{v}{c}\cos{\phi})$, with $f$ representing the scattering frequency.

3) Change in photon energy and direction for scattering events are significantly greater for head-on photons, and vice versa, as the electron distribution isn't isotropic in the BH rest frame, by $\tan{\theta^\prime} = \frac{u\sin{\theta}}{\gamma(u\cos{\theta} + v)}$, where $\gamma$ is the electron's Lorentz factor, $u$ the electron velocity, and $\theta$ is the respective electron angle of travel.

	These effects, taken together, typically result in emissions that are stronger when initially emitted, but have somewhat lower scattering luminosities, as most photons will be beamed in a similar direction to the bulk flow, so will scatter less frequently and with less change in angle.  This can be seen in Figure 11, which depicts single zone trials.
	
\begin{figure}
\includegraphics[scale=.35]{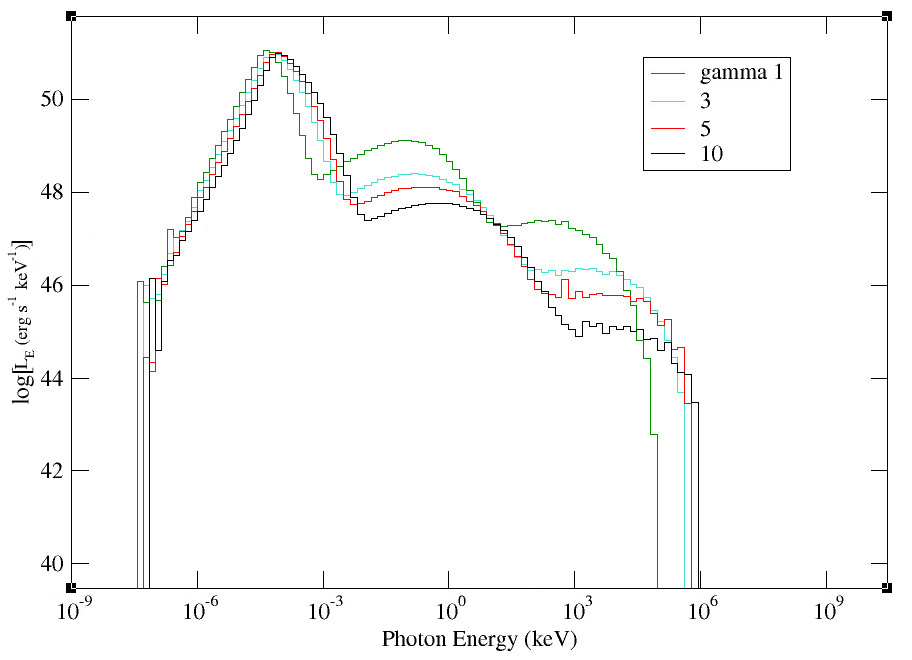}
\caption{A single zone trial to illustrate the effects of the suite of velocity modifications.  The general effect of higher Lorentz factor (gamma) can be seen in the boosting to higher energies, but the loss of scattering frequency, as photons are beamed in the direction of electron travel.}
\end{figure}

\end{document}